\documentclass[aps,pra,showpacs,floatfix,twocolumn,superscriptaddress]{revtex4}
\usepackage[english]{babel}
\usepackage[T1]{fontenc}
\usepackage{amsmath,amsfonts,amssymb,float,graphicx,color,times,bbm,stmaryrd,pxfonts}
\usepackage[bookmarks=false]{hyperref}
\newcommand{\id}{{\sf 1 \hspace{-0.3ex} \rule{0.1ex}{1.52ex}\rule[-.01ex]{0.3ex}{0.1ex}}}
\renewcommand{\L}[2]{\Lambda_{
\mbox{{\tiny $#2$}}}^{\mbox{{\tiny (#1)}}}}
\renewcommand{\O}[2]{\Omega_{
\mbox{{\tiny $#2$}}
}^{\mbox{{\tiny (#1)}}}}
\newcommand{\cL}[2]{\Lambda_{
\mbox{{\tiny $#2$}}}^{\mbox{{\tiny (#1)}{\footnotesize *}}}}
\newcommand{\cO}[2]{\Omega_{
\mbox{{\tiny $#2$}}
}^{\mbox{{\tiny (#1)}{\footnotesize *}}}}
 
\newcommand{\Ph}[2]{\Phi_{
\mbox{{\tiny $#2$}}
}^{\mbox{{\tiny (#1)}}}}
\newcommand{\cPh}[2]{\Phi_{
\mbox{{\tiny $#2$}}
}^{\mbox{{\tiny (#1)}{\footnotesize *}}}}

\begin{document}
\author{J. P. Santos}
\affiliation{Centre for Theoretical Atomic, Molecular and Optical Physics,
School of Mathematics and Physics, Queen's University Belfast, BT7 1NN, United Kingdom}
\affiliation{Centro de Ci\^encias Naturais e Humanas, Universidade Federal do ABC, Santo Andr\'e, 09210-170 S\~ao Paulo, Brazil}
\title{
Master equation for dissipative interacting qubits in a common environment}
\author{F. L. Semi\~ao}
\affiliation{Centro de Ci\^encias Naturais e Humanas, Universidade Federal do ABC, Santo Andr\'e, 09210-170 S\~ao Paulo, Brazil}
\begin{abstract}
In this paper, we derive a microscopic master equation for a pair of $XY$-coupled two-level systems interacting with the same memoryless reservoir. In particular, we apply this master equation to the case of a pair of two-level atoms in free space where we can clearly contrast the predictions made with the microscopic master equation obtained here and the phenomenological approaches where the atom-atom coupling is included just \textit{a posteriori}, i.e, not taking into account in the derivation of the open system equation of motion. We show, for instance, that the phenomenological approach fails completely in the assessment of the role played by the symmetric and antisymmetric decay channels. As a consequence, the predictions related to collective effects such as superradiance, for instance, are misleading in the phenomenological approach. We also obtain the fluorescence spectrum using the microscopic model developed here. 
\end{abstract}
\maketitle

\section{Introduction}
%
The dynamics of open quantum systems have been studied extensively in the fields of quantum optics \cite{carmichael}, quantum information \cite{reina}, and more recently in quantum biophysics \cite{biodec}. The main goal is to describe the non-unitary behavior resulting from the fact that the system is not closed. An usual and useful approach is the use of master equations for the system density operator \cite{master1}. These master equations can describe memoryless Markovian evolutions as well as non-Markovian evolutions \cite{master2}. Usually, such master equations are obtained by considering a microscopic model for the interaction of the system under study and the environment, and  tracing out the environment variables in some exact or, most of the times, perturbative treatment \cite{carmichael}.

Usually, the presence of interactions among parts of the system or its subsystems is not taken into account in the derivation of the master equation. It is not uncommon to find examples of such a \textit{phenomenological} approach where one consider the response of a system to the environment to be exactly the same regardless of whether it is coupled or not to another quantum system. As an illustrative example, let us consider the celebrated Jaynes-Cummings model describing the interaction of a single-mode quantized electromagnetic field in a cavity and a two-level atom in the rotating wave approximation \cite{jcm}. When considering cavity losses at a decay rate $\kappa$, one usually describes the open system by using a master equation in the form
\begin{eqnarray}\label{jcm}
\partial_t{\rho}=\frac{1}{i\hbar}[H_{JC},\rho]+\kappa(a\rho a^\dag-a^\dag a\rho-\rho a^\dag a).
\end{eqnarray}
The first term on right-hand side of (\ref{jcm}) is the coherent part of the evolution, which is generated by the Jaynes-Cummings Hamiltonian $H_{JC}$. More details about this Hamiltonian can be found in \cite{jcm}. The second term, known as Liouvillian, accounts for the losses. The potential problem of using (\ref{jcm}) resides in the fact that the derivation of the Liouvillian was realized in another different microscopic model, and not the Jaynes-Cummings plus environment. Actually, this Liouvillian is deduced for a cavity mode losing photons to the vacuum environment \textit{without} the presence of the atom. For this reason, the use of (\ref{jcm}) is a phenomenological approach. A microscopic derivation of the master equation for the Jaynes-Cummings model with cavity losses is found in \cite{sabrina}.

Here, we will be interested in a system formed by a pair of XY-coupled qubits which, due to interaction with the environment, are subject to energy relaxation. We will be particularly interested in \textit{collective effects} resulting from the coupling of each subsystem to the same reservoir or environment. This situation appears, for instance, in the Dicke superradiance phenomenon involving atoms and the quantized electromagnetic field \cite{Dicke}. We will be concerned with coupled qubits because this kind of system is present in most of the modern applications of quantum mechanics \cite{qi}. Furthermore, experimental advances in the control of single quantum systems has achieved a maturity such that precise control over qubit interactions is achieved in many different setups \cite{nobel}. Consequently, acquiring a thoroughly understanding and an accurate description of the open system dynamics of coupled qubits is a timely problem. Additionally, coupled qubits also appear in very complex environments such as in naturally occurring coupled chromofores supporting single excitons in pigment-protein complexes  \cite{Thorwart2009234,1367-2630-11-8-085001, transfer} or in conjugated polymer samples \cite{poly}.

In this contribution, we perform the microscopic derivation of a master equation taking into account, from the start, the collective effects of a common reservoir and the XY interaction between the qubits. We then throughly compare results predicted by the obtained generalized master equation to that predicted using phenomenological models. This work is organized as follows. In Section \ref{derivation}, we do the microscopic derivation of the master equation for two XY-coupled qubits in interaction with a common memoryless reservoir. In Section \ref{atomic}, we particularize it for the case of two atoms in free space which are also subjected to an externally controlled XY interaction. For this system, we study the populations in the collective basis in Section \ref{P}, and we obtain the fluorescence spectrum in Section \ref{F}. In Section \ref{S}, we summarize our results. In order to clarify the exposition of the ideas, we reserved Appendix \ref{app1} to present the most lengthy expressions and also the results already known in the literature. 
\section{Generalized master equation} \label{derivation}
In this work, we will be interested in the study two qubits coupled by means of a XY interaction. The Hamiltonian describing the closed system dynamics reads
\begin{eqnarray}\label{f1}
H_S=\frac{\hbar\omega_1}{2}\sigma^z_1+\frac{\hbar\omega_2}{2}\sigma^z_2-\hbar g(\sigma^x_1\sigma^x_2+\sigma^y_1\sigma^y_2),
\end{eqnarray}
where $\omega_i$ is the angular frequency of qubit $i$, which undergoes transitions $|g\rangle_i\rightleftharpoons|e\rangle_i$, the $\sigma$'s are the usual Pauli matrices, and $g$ is the XY coupling constant. By defining $J=2g$ and using $\sigma_i^+=|g\rangle_i\langle e|$ ($\sigma_i^-=[\sigma_i^+]^\dag$), we can rewrite (\ref{f1}) as
\begin{eqnarray}\label{f2}
H_S=\frac{\hbar\omega_1}{2}\sigma^z_1+\frac{\hbar\omega_2}{2}\sigma^z_2-\hbar J(\sigma^+_1\sigma^-_2+\sigma^-_1\sigma^+_2),
\end{eqnarray}
which is the form used from now on. This form of coupling between the qubits is also known as dipole-dipole coupling in the electromagnetic context. 

The simplest phenomenological approach for this system consists of using a master equation of the form 
\begin{align}
\label{ME1}
\partial_t{\rho}=\frac{1}{i\hbar}[H_S^{(1)},\rho]
+\mathcal{L}_1(\rho)+\mathcal{L}_2(\rho),
\end{align}
where Hamiltonian $H_S^{(1)}$ contains the XY interaction between the qubits and a Lamb shift term for each qubit due to individual coupling with a reservoir. Liouvillian $\mathcal{L}_i(\rho)$ contains only operators acting on qubit $i$. The actual form of $H_S^{(1)}$ and $\mathcal{L}_i(\rho)$ are given in Appendix \ref{app1}, equations (\ref{s1}) and (\ref{l1}), respectively. Each Liouvillian $\mathcal{L}_i(\rho)$ on the right-hand side of (\ref{ME1}) has been derived microscopically in the framework of a \textit{different} physical problem where the system is composed by just one qubit in contact with its reservoir. Then, predictions made with (\ref{ME1}) might deviate greatly from the observations for not taking the qubit-qubit coupling into consideration in its derivation. Master equation (\ref{ME1}) is also expected to fail when collective effects arising from coupling to the same reservoir are involved.

A more elaborated phenomenological approach could make use of a master equation of the form
\begin{align}\label{ME2}
\partial_t{\rho}=\frac{1}{i\hbar}[H_S^{(2)},\rho]
+\mathcal{L}_{12}(\rho)+\mathcal{L}_{12}^\dag(\rho)
+\sum_{i=1}^2\mathcal{L}_i(\rho),
\end{align}
where the Hamiltonian $H_S^{(2)}$ contains individual and collective Lamb shifts, being the latter a kind of coherent interaction between the qubits induced by the common reservoir. In this case, the Liouvillians were deduced with decoupled qubits ($J=0$) in a single common or shared bosonic environment \cite{ficek}. This shared reservoir induced a collective Liouvillian $\mathcal{L}_{12}$ acting on the state of both qubits, besides an already mentioned coherent interaction entering $H_S^{(2)}$. Expressions for $H_S^{(2)}$ and  $\mathcal{L}_{12}(\rho)$ can be found in the Appendix \ref{app1}, in equations (\ref{hs2}) and (\ref{l12}), respectively. Also, in order to obtain (\ref{ME2}) in this concise form, one makes the approximation $\L{-}{12\omega_i}\approx\L{-}{12\omega_0}$ and $\O{-}{12\omega_i}\approx\O{-}{12\omega_0}$, where $\L{-}{12\mu}$ and $\O{-}{12\mu}$ are defined in (\ref{l12}) and (\ref{Oij}), respectively, and $\omega_0=(\omega_1+\omega_2)/2$ \cite{ficek}.
This is possible when the difference between the atomic frequencies $|\omega_1-\omega_2|$ are much smaller than the average atomic frequency $\omega_0$ \cite{ficek}. Clearly, this is the case when dealing with qubits of the same nature having comparable transition frequencies. This master equation can be a good approximation only if the externally induced XY interaction between the qubits is weak enough. If this is not the case, the equilibrium state of the composite system is not guaranteed to be equal of that of the non interacting system \cite{kubo}. It is to properly deal with strongly coupled qubits in the presence of a shared memoryless reservoir that in this work we derive a master equation from a microscopic model that takes into account the qubit-qubit coupling.  

Let us then consider a system $\mathcal{S}$ formed by a pair of XY-coupled qubits, and a reservoir $\mathcal{R}$ consisting of a collection of independent harmonic oscillators. We will denote $\rho_T(t)$ the density operator for $\mathcal{S}\otimes\mathcal{R}$, and define the reduced density operator $\rho(t)$ obtained by tracing out the reservoir $\rho(t)={\rm tr}_{\mathcal{R}}[\rho_T(t)]$. The Schrödinger equation for $\rho_T(t)$ reads
\begin{eqnarray}
\label{ME}
\partial_t{\rho}_T=\frac{1}{i\hbar}[H,\rho_T],
\end{eqnarray}
where
$\label{HAM}
H=H_S+H_R+H_{SR},
$
with $H_S$ given by (\ref{f2}), and
\begin{eqnarray}
\label{reservoir}
H_R&=&\sum_{{\bf k},\lambda}\omega_ka^{+}_{{\bf k},\lambda}a^{}_{{\bf k},\lambda},
\\
H_{SR}&=&\sum_{{\bf k},\lambda}(
\sigma_1^{+}\kappa^{}_{1;{\bf k},\lambda}+
\sigma_2^{+}\kappa^{}_{2;{\bf k},\lambda})a^{}_{{\bf k},\lambda}
+{\rm H.c.},\label{sr}
\end{eqnarray}
where $\omega_k$ is the angular frequency of oscillator ${\bf k}$, $\kappa^{}_{i;{\bf k},\lambda}$ is a coupling constant (for the moment unspecified), and $\lambda$ may account for polarization in the case of the electromagnetic vacuum. By moving (\ref{ME}) to the interaction picture $[\widetilde{(\diamond)}=e^{\,i/(H_S+H_R)t}\,(\diamond)\,e^{-i/(H_S+H_R)t}]$, formally integrating the result, and iterating it one time, one obtains \cite{carmichael}
\begin{eqnarray}\label{antes}
\partial_t{\widetilde{\rho}}_T
&=&\widetilde{\rho}_T(0)+\frac{1}{i\hbar}
[\widetilde{H}_{SR}(t),
\widetilde{\rho}_T(0)]
 \nonumber\\ &&-\frac{1}{\hbar^2}
\int_0^t{\rm d}t'[\widetilde{H}_{SR}(t)
,[\widetilde{H}_{SR}(t')
,\widetilde{\rho_T}(t')]].
\end{eqnarray} 
Assuming no correlations between $\mathcal{S}$ and $\mathcal{R}$ at $t=0$, we can write
$
\rho_T(0)=\rho(0)\otimes\rho_\mathcal{R}(0),
$
where $\rho_\mathcal{R}(0)$ is the initial reservoir density operator. The state of the reservoir will be chosen to be equilibrium state of independent oscillators at temperature $T=0{\rm K}$ (vacuum). Now, by tracing out the reservoir in (\ref{antes}), and performing the Born and Markov approximations \cite{carmichael,ficek} one gets
\begin{eqnarray}\label{antes2}
\partial_t{\widetilde{\rho}}=
-\frac{1}{\hbar^2}\int_0^t{\rm d}t'
{\rm tr}\Big\{[
\widetilde{H}_{SR}(t)
,[\widetilde{H}_{SR}(t'),
\widetilde{\rho}(t)\rho_{\mathcal{R}}(0)]
]\Big\}.
\end{eqnarray}
Finally, by writing the transformed interaction Hamiltonian as
$\widetilde{H}_{SR}=\hbar\sum_i 
\widetilde{s}_i\widetilde{\Gamma}_i,
$
where $s_i$ acts on $\mathcal{S}$ and $\Gamma_i$ acts on $\mathcal{R}$, and changing variables ($\tau=t-t'$), we may rewrite  (\ref{antes2}) as
\begin{widetext}
\begin{eqnarray} \label{1.18carmichael}
\partial_t{\widetilde{\rho}}&=&\sum_{ij}\int_0^t{\rm d}
\tau\{
[\widetilde{s}_j(t-\tau)\widetilde{\rho}(t)
\widetilde{s}_i(t)
-\widetilde{s}_i(t)\widetilde{s}_j(t-\tau)
\widetilde{\rho}(t)
]
\langle\widetilde{\Gamma}_i(t)\widetilde{\Gamma}_j(t-\tau)\rangle_{\mathcal R}
+[
\widetilde{s}_i(t)\widetilde{\rho}(t)\widetilde{s}_j(t-\tau)
-\widetilde{\rho}(t)\widetilde{s}_j(t-\tau)\widetilde{s}_i(t)
]\langle\widetilde{\Gamma}_j(t-\tau)\widetilde{\Gamma}_i(t)\rangle_{\mathcal R}
\}.\nonumber\\
\end{eqnarray}
\end{widetext}
Equation (\ref{1.18carmichael}) is quite general, and it sets the ground for working out different master equations depending on the microscopic model at hand. In our case, following (\ref{sr}), we must have
$
s_1=\sigma_1^{-}, s_2=\sigma_1^{+}
, 
s_3=\sigma_2^{-}, s_4=\sigma_2^{+},
$ and
$
\Gamma_1=\sum_{{\bf k},\lambda}\kappa^*_{1;{\bf k},\lambda}a^{+}_{{\bf k},\lambda}, 
\Gamma_2=\sum_{{\bf k},\lambda}\kappa_{1;{\bf k},\lambda}a_{{\bf k},\lambda},
\Gamma_3=\sum_{{\bf k},\lambda}\kappa^*_{2;{\bf k},\lambda}a^{+}_{{\bf k},\lambda},  
\Gamma_4=\sum_{{\bf k},\lambda}\kappa_{2;{\bf k},\lambda}a_{{\bf k},\lambda}.
$
For $H_S$ given by (\ref{f2}), which contains the XY interaction, and $H_R$ given by (\ref{reservoir}), which is the common reservoir, we find

\begin{eqnarray}\label{se}
\widetilde{s}_1(t)&=&\theta_1^*(t)\sigma_1^{-}+\theta_2^*(t)\sigma_1^z\sigma_2^-,
\\
\widetilde{s}_2(t)&=&\theta_1(t)\sigma_1^{+}+\theta_2(t)\sigma_1^z\sigma_2^+,
\\
\widetilde{s}_3(t)&=&\phi_1^*(t)\sigma_2^{-}+\theta_2^*(t)\sigma_1^-\sigma_2^z,
\\
\widetilde{s}_4(t)&=&\phi_1(t)\sigma_2^{+}+\theta_2(t)\sigma_1^+\sigma_2^z,
\end{eqnarray}
where
\begin{eqnarray}
\theta_1(t)&=&\delta e^{-i\beta t}+\gamma e^{i\alpha t}
\\
\theta_2(t)&=&\frac{J}{\Delta}(e^{i\alpha t}-e^{i\beta t})
\\
\phi_1(t)&=&\gamma e^{-i\beta t}+\delta e^{i\alpha t}
\end{eqnarray}
with
\begin{eqnarray}\label{af}
\Delta&=&\sqrt{4J^2+(\omega_1-\omega_2)^2},\\ \label{afa}
\label{alpha}\alpha&=&(\Delta+\omega_1+\omega_2)/2, \\ \label{afb}
\label{beta}\beta&=&(\Delta-\omega_1-\omega_2)/2, \\ \label{afc}
\gamma&=&(\Delta+\omega_1-\omega_2)/(2\Delta), \\ \label{afd}
\delta&=&(\Delta-\omega_1+\omega_2)/(2\Delta),
\end{eqnarray}
and, for the reservoir operators, we find
$
\widetilde{\Gamma}_1=\sum_{{\bf k},\lambda}\kappa_{1;{\bf k},\lambda}^*a_{{\bf k},\lambda}^+e^{i\omega_kt}, 
\widetilde{\Gamma}_2=\sum_{{\bf k},\lambda}\kappa_{1;{\bf k},\lambda}a_{{\bf k},\lambda}e^{-i\omega_kt},
\widetilde{\Gamma}_3=\sum_{{\bf k},\lambda}\kappa_{2;{\bf k},\lambda}^*a_{{\bf k},\lambda}^+e^{i\omega_kt}, 
\widetilde{\Gamma}_4=\sum_{{\bf k},\lambda}\kappa_{2;{\bf k},\lambda}a_{{\bf k},\lambda}e^{-i\omega_kt}.
$

It is already possible to notice that the presence of the XY interaction between the qubits will bring new terms not present in (\ref{ME2}). For example, let us consider the time evolution of $\widetilde{s}_1(0)=\sigma_1^-$. According to (\ref{se}), the $J-$coupling causes the appearance of a new term which is proportional to $\sigma_1^z\sigma_2^-$, absent in (\ref{ME2}). From (\ref{af}) to (\ref{afd}), one can also see that the frequencies and coefficients appearance in (\ref{1.18carmichael}) strongly depend on $J$. In next sections, we will see how these new terms influence the dynamics and spectrum of the system. 

By considering a memoryless reservoir, the integrand in (\ref{1.18carmichael}) is not negligible only for very short times compared with time scale for the evolution of $\widetilde{\rho}$. Then, we can extend the upper integration limit to infinity, and perform the integrations in $\tau$ using
\begin{eqnarray}\label{cauchy}
\int_0^{\infty}\exp(\pm i \xi\tau) d\tau=
\pi\delta(\xi)\pm i\mathcal{P}\frac{1}{\xi},
\end{eqnarray}
where $\mathcal{P}$ denotes the Cauchy principal value. In our case, $\xi$ stands for $\omega_k\pm\alpha$ and $\omega_k\pm\beta$.
We also make the usual transition to the continuum \cite{carmichael}, what allows us to integrate the reservoir variables in ${{\bf k}}$. After all these integrations, (\ref{1.18carmichael}) becomes 
\begin{widetext}
\begin{eqnarray}\label{fi}
\begin{split}
\partial_t{\rho}&=\frac{1}{i\hbar}[\rho,H_S]
+\sum_{ij}\Big(
\mathcal{A}_{ij} \ \sigma_i^{-}\rho\sigma_j^{+} 
+\mathcal{B}_{ij} \ \sigma_i^{+}\sigma_j^{-}\rho 
+\mathcal{B}_{ij}^* \ \rho\sigma_i^{-}\sigma_j^{+}\Big)\\
&+\sum_{i\neq j}\Big(
\mathcal{C}_{ij}\sigma_i^{z}\sigma_j^{-}\rho\sigma_i^{+} 
+\mathcal{D}_{ij}\sigma_i^{z}\sigma_j^{-}\rho\sigma_j^{+}
+\mathcal{E}_{ij}\sigma_i^{+}\sigma_i^{-}\sigma_j^{z}\rho 
+\mathcal{F}_{ij}\sigma_i^{-}\sigma_i^{+}\sigma_j^{z}\rho+ {\rm H.c.}\Big),
\end{split}
\end{eqnarray}
\end{widetext}
where $H_S$ is given by (\ref{f2}) and the coefficients $\mathcal{A}_{ij}$, $\mathcal{B}_{ij}$, $\mathcal{C}_{ij}$, $\mathcal{D}_{ij}$, $\mathcal{E}_{ij}$ and $\mathcal{F}_{ij}$ are defined in (\ref{coef}) in Appendix \ref{app1}. We presented (\ref{fi}) already in the original Schrödinger picture, i.e., we have performed $(\diamond)=e^{-i/\hbar(H_S+H_R)t}\,\widetilde{(\diamond)}\,e^{\,i/\hbar(H_S+H_R)t}$. This is the master equation microscopically deduced for two qubits coupled through a XY interaction in the presence of a common bosonic reservoir at $T=0{\rm K}$. 

Master equation (\ref{fi}) contains the local decays $\mathcal{L}_{i}$  present in (\ref{ME1}) and (\ref{ME2}) and the collective decays $\mathcal{L}_{12}$ present only in (\ref{ME2}). However, these terms have coefficients which now depend on $J$. All coherent Lamb-shifts, including an induced dipole-dipole coupling, are present in the imaginary part of the coefficients in (\ref{fi}). One can see that (\ref{fi}) contains phase elements, through the presence of the operator $\sigma_i^z$, something missing in (\ref{ME1}) and (\ref{ME2}). These terms are accompanied by coefficients $\mathcal{C}_{ij}$, $\mathcal{D}_{ij}$, $\mathcal{E}_{ij}$ and $\mathcal{F}_{ij}$ which, as expected, goes to zero at the limit of decoupled qubits, but affect the system dynamics for finite $J$. 

In order to objectively compare the results predicted by (\ref{fi}) with that predicted (\ref{ME1}) or (\ref{ME2}), we know work within an specific example. In next section, we will consider two XY-coupled atoms interacting with the electromagnetic vacuum. We will suppose that the XY interaction is externally controlled by some mechanism such a the one presented in \cite{zheng}. The precise mechanism leading to this qubit-qubit coupling will not concern us here.
\section{Two coupled atoms in the Electromagnetic Vacuum}\label{atomic}
We now particularize to the case in which the reservoir described by (\ref{reservoir}) is the electromagnetic vacuum. In this case, atom $i$ is coupled to the reservoir through \cite{carmichael}
\begin{eqnarray}\label{kappa}
\kappa_{i;{\bf k},\lambda}^{}=
-ie^{i{\bf k}\cdot{\bf r}_i}\sqrt{\frac{\omega_k}{2\hbar\epsilon_0V}}
{\bf e}_{{\bf k},\lambda}\cdot{\bf d}_{i},
\end{eqnarray}
where ${\bf d}_{i}$ is the electric dipole of atom $i$, and the summation in (\ref{reservoir}) extends over all free space electromagnetic field modes with wavevectors  ${\bf k}$ (angular frequency $\omega_k$) and polarizations $\lambda$. The atom $i$ is positioned at ${\bf r}_i$ and $V$ is the quantization volume. In order to change to a continuum, we will need a density of states $g({\bf k})$ for each polarization $\lambda$  \cite{carmichael}
\begin{eqnarray}\label{g}
g({\bf k}){\rm d}^3k=\frac{\omega^2V}{8\pi^3c^3}d\omega_k\sin\theta{\rm d}\theta{\rm d}\phi,
\end{eqnarray}
where we dropped the index $k$ in $\omega_k$.  Using (\ref{kappa}) and (\ref{g}), and assuming parallel atomic dipoles, we  calculated (\ref{Oij}) and (\ref{Lij}) obtaining
\begin{widetext}
\begin{eqnarray}\label{Oiju}
\O{$\pm$}{ij\mu}=
\left\{
\begin{array}{l}
\mp\frac{3}{2}\frac{|{\bf d}_i||{\bf d}_j|\mu^3}{3\pi\epsilon_0\hbar c^3}
\Big[
\sin^2(\Theta)\sin(\chi_\mu)/\chi_\mu
+(1-3\cos^2\Theta)\left(
\frac{\cos\chi_\mu}{\chi_\mu^2}-
\frac{\sin\chi_\mu}{\chi_\mu^3}
\right)\Big],\hspace{1.06cm}{\rm if} \ \  \mu<0,\\
 \ 0, \ \hspace{9.9cm}   {\rm if}  \ \ \mu>0,
\end{array}
\right.
\end{eqnarray}
\end{widetext}
and
\begin{widetext}
\begin{eqnarray}\label{Liju}
\L{$\pm$}{ij\mu}=
\left\{
\begin{array}{l}
\mp\frac{3}{8}\frac{|{\bf d}_i||{\bf d}_j||\mu|^3}{3\pi\epsilon_0\hbar c^3}
\left\{
(1-3\cos^3\Theta)
\left(
\frac{\sin|\chi_\mu|}{\chi_\mu^2}
+\frac{\cos\chi_\mu}{|\chi_\mu|^3}\pm\frac{2}{\pi}
\left[
\frac{F_2(|\chi_\mu|)}{\chi_\mu^2}+
\frac{F_1(|\chi_\mu|)}{|\chi_\mu|^3}\right]
\right)
\right. 
\left.
\pm\sin^2\Theta
\left(
\frac{\cos\chi_\mu}{|\chi_\mu|}
+
\frac{2}{\pi}
\left[
\frac{1}{\chi_\mu^2}
-\frac{F_1(|\chi_\mu|)}{|\chi_\mu|}
\right]
\right)
\right\} \hspace{0.5cm}, {\rm if} \ \   \mu>0,
\nonumber\\
\mp\frac{3}{8}\frac{|{\bf d}_i||{\bf d}_j||\mu|^3}{3\pi\epsilon_0\hbar c^3}
\left\{
(1-3\cos^3\Theta)
\left(
\frac{\sin|\chi_\mu|}{\chi_\mu^2}
+\frac{\cos\chi_\mu}{|\chi_\mu|^3}\mp\frac{2}{\pi}
\left[
\frac{F_2(|\chi_\mu|)}{\chi_\mu^2}+
\frac{F_1(|\chi_\mu|)}{|\chi_\mu|^3}\right]
\right)
\right.
\left.
\mp\sin^2\Theta
\left[
\frac{\cos\chi_\mu}{|\chi_\mu|}
+
\frac{2}{\pi}
\left[\frac{1}{\chi_\mu^2}
-\frac{F_1(|\chi_\mu|)}{|\chi_\mu|}\right]
\right]\right\}\hspace{0.5cm}, {\rm if} \ \   \mu<0.
\end{array}
\right. \\
\end{eqnarray}
\end{widetext}
In these expressions, $\mu$ can either be $\alpha$ or $\beta$ given by (\ref{alpha}) and (\ref{beta}), respectively, and $\chi_\mu\equiv\mu r_{12}/c$, where $r_{12}$ is the atom-atom distance defined as $r_{12}=|{\bf r}_1-{\bf r}_2|$.  $\Theta$ is the angle between 
the electric dipole of each atom (assumed to be parallel) and ${\bf r}_{12}={\bf r}_1-{\bf r}_2$. Also, 
\begin{eqnarray}
F_1(\xi)&=&\sin(\xi){\rm Ci}(\xi)-\cos(\xi){\rm Si}(\xi), \\
F_2(\xi)&=&-\sin(\xi){\rm Si}(\xi)-\cos(\xi){\rm Ci}(\xi),
\end{eqnarray}
where ${\rm Si}[{\rm Ci}]$ is the sine [cosine] integral function. 

The terms $\O{$\pm$}{i\mu}$ can be calculated by using  $\O{$\pm$}{i\mu}=\O{$\pm$}{ii\mu}$ (\ref{def1}). However, the individual lamb-shift terms $\L{$\pm$}{i\mu}$ evaluated with $ \L{$\pm$}{i\mu}=\L{$\pm$}{ii\mu}$ (\ref{def2}) are, in the scope of this nonrelativistic theory, mathematically divergent.
It is possible to give them a finite value through introducing a frequency cutoff $\omega_c$ and including the counter rotating terms. The result is \cite{PhysRevA.7.1195,PowerZienau}
\begin{eqnarray}
\L{$\pm$}{i\mu}=\frac{1}{2\pi}\frac{|{\bf d}_i|^2\mu^3}{3\pi c^3\hbar \epsilon_0}
\ln\left[\left|\frac{\omega_c}{|\mu|}-1\right|
\left(\frac{\omega_c}{|\mu|}+1\right)\right].
\end{eqnarray}
From the practical side though, there is no much worry because these Lamb-shifts can be seen as renormalized atomic frequencies which are, as we all know, finite.

If we consider that $\omega_1=\omega_2$ and ${\bf d}_1={\bf d}_2$, some quantities become identical to both atoms,  specially $\omega_0=\omega_1=\omega_2$, which allows us to define $\L{$\pm$}{}\equiv\L{$\pm$}{1}=\L{$\pm$}{2}$ and $\O{$\pm$}{}\equiv\O{$\pm$}{1}\equiv\O{$\pm$}{2}$, where  $\L{$\pm$}{i}$ and $\O{$\pm$}{i}$ are given by (\ref{def3}) and (\ref{def4}), respectively. By using (\ref{coef}) and (\ref{fi}), we then obtain the following master equation describing two identical atoms, with parallel dipole moments, and coupled by means of an external XY interaction
\begin{widetext}
\begin{eqnarray}\label{ME4}
 \begin{split}
\partial_t\rho=&\frac{1}{i\hbar}[H_S^{(3)},\rho]
+\frac{\O{+}{}}{2}\sum_{i=1}^2\big(2
\sigma_i^{-}\rho\sigma_i^{+}-\{\sigma_i^{+}\sigma_i^{-},\rho\}\big)
+\sum_{{i,j}=1\atop i\neq j}^2\Big\{\frac{1}{2}(
\O{+}{12}-\O{-}{})(2\sigma_j^{-}\rho\sigma_i^{+}-\{\sigma_i^{+}\sigma_j^{-},\rho\}\big)
\\ & 
+\frac{\O{-}{12}}{2}
\big(
\sigma_i^z\sigma_j^-\rho\sigma_j^+
+\sigma_i^-\rho\sigma_i^+\sigma_j^z
-\{\sigma_i^+\sigma_i^-\sigma_j^z,\rho\}\big)
+\frac{i\L{+}{12}}{2}
\big(
\sigma_i^z\sigma_j^-\rho\sigma_j^+
-\sigma_j^-\rho\sigma_i^z\sigma_j^+
-[\sigma_i^+\sigma_i^-\sigma_j^z,\rho]\big)
\\ & 
+\frac{\O{-}{}}{2}
\big(
\sigma_i^z\sigma_j^-\rho\sigma_i^+
+\sigma_i^-\rho\sigma_i^z\sigma_j^+
+2\sigma_i^{-}\rho\sigma_j^{+}\big)
+\frac{i\L{+}{}}{2}
\big(
\sigma_i^z\sigma_j^-\rho\sigma_i^+
-\sigma_i^-\rho\sigma_i^z\sigma_j^+\big)\Big\},
\end{split}
\end{eqnarray}
\end{widetext}
where Hamiltonian $H_S^{(3)}$ is given by
\begin{eqnarray}\label{HS3}
H_S^{(3)}&=&\frac{1}{2}\hbar(\omega_0+\L{-}{}/2)
(\sigma_1^{z}+\sigma_2^{z})\nonumber\\ &&-\hbar\bigg[J+\frac{1}{2}(
\L{+}{}
-\L{-}{12}
)\bigg](\sigma_1^+\sigma_2^-+\sigma_1^-\sigma_2^+),
\end{eqnarray}
with $\L{$\pm$}{12}$ and $\O{$\pm$}{12}$ are given by (\ref{def5}) and (\ref{def6}). It is this master equation which will now be studied in detail. We will see the role played by $J$ in the dynamics and fluorescence spectrum of the system, contrasting the results obtained with (\ref{ME4}) and the phenomenological models (\ref{ME1}) and (\ref{ME2}).

It is important to emphasize that master equation (\ref{ME4}) reduces to (\ref{ME2}) when $J\rightarrow 0$, and (\ref{ME2}) reduces to (\ref{ME1}) when $r_{12}\gg\lambda$, where $\lambda=2\pi c/\omega_0$. Therefore, when the XY coupling constant $g$ in (\ref{f1}) or $J$ in (\ref{f2}) is weak enough, master equation (\ref{ME2}) is expected to become a good approximation to the generalised master equation (\ref{ME4}) which was obtained here. At the same token, if $J$ is weak enough and the atoms are sufficiently far apart, master equation (\ref{ME1}) is likely to be a good approximation for (\ref{ME2}) and (\ref{ME4}). In next sections, we will compare predictions made with these models, and we will see that for moderate $J$ the differences are big enough to be experimentally testable.

\section{Populations} \label{P}
\newcommand{\e}{\varepsilon}
\newcommand{\g}{\varg}
We begin this section defining the collective states, also known as ``molecular'' eigenstates or ``exciton'' basis. They are given by

\begin{eqnarray}\label{collective}
 \begin{split}
|\e\rangle =&|ee\rangle,\\
|\g\rangle =&|gg\rangle,\\
|s\rangle =&\frac{|eg\rangle+|ge\rangle}{\sqrt{2}},\\
|a\rangle =&\frac{|eg\rangle-|ge\rangle}{\sqrt{2}}.
\end{split}
\end{eqnarray}
These are the eigenstates of the XY Hamiltonians (\ref{f1}) and (\ref{f2}), and they provide a natural basis for discussing symmetries and spectrum of multiatom systems \cite{ficek}. We will now then rewrite the master equation (\ref{ME4}) in this basis, and to achieve this, it is convenient to define collective operators $A_{ij}=|i\rangle\langle j|$, where $i,j=\{\e,\g,a,s\}$. These collective operators are related to the individual atomic ones by
\begin{eqnarray}\label{trans}
\begin{split}
\sigma_1^{+}&=&\sigma_1^{+}\otimes\id_2=\frac{1}{\sqrt{2}}(A_{\e s}-A_{\e a}+A_{s\g}+A_{a\g}), \\
\sigma_2^{+}&=&\id_1\otimes\sigma_2^{+}=\frac{1}{\sqrt{2}}(A_{\e s}+A_{\e a}+A_{s\g}-A_{a\g}).
\end{split}
\end{eqnarray}
Now, we simply substitute (\ref{trans}) in (\ref{ME4}) to obtain
\begin{eqnarray}\label{ME5}
\partial_t{\rho}=\frac{1}{i\hbar}[H_{S}^{(4)},\rho]+\mathcal{L}_{\,a}\rho+\mathcal{L}_{\,s}\rho,
\end{eqnarray}
 where $H_S^{(4)}$ is given by
\begin{eqnarray}\label{HS4t}
H_S^{(4)}=
\hbar\omega'(A_{ee}-A_{gg})
+\hbar J^{(-)} A_{aa}-\hbar J^{(+)} A_{ss},
\end{eqnarray}
with
\begin{eqnarray}\label{cf}
J^{(\pm)}=J+\frac{1}{2}(\L{+}{}-\L{-}{12}\pm 2\L{+}{12}),
\end{eqnarray}
and
\begin{eqnarray}\label{wf}
\omega'=\omega_0+\frac{\L{-}{}+\L{+}{12}}{2}.
\end{eqnarray}
In (\ref{ME5}), we also used $\mathcal{L}_{\,a}\rho$ and $\mathcal{L}_{\,s}\rho$ defined as
\begin{widetext}
\begin{eqnarray}
\begin{split}
\mathcal{L}_{\,a}\rho=&
-\frac{1}{2}(\O{+}{}-\O{+}{12})
\big[(A_{aa}+A_{\e\e})\rho+\rho(A_{aa}+A_{\e\e})
-2(A_{a\e}-A_{\g a})\rho(A_{\e a}-A_{a\g})\big]
\\&
-\frac{1}{2}(\O{-}{}-\O{-}{12})
\big[(A_{aa}-A_{\e\e})\rho+\rho(A_{aa}-A_{\e\e})
-2(A_{\g a}\rho A_{a\g}-A_{ae}\rho A_{\e a})\big]
+i(\L{+}{}-\L{+}{12})(A_{a\e}\rho A_{a\g}-A_{\g a}\rho A_{\e a}),
\end{split}
\end{eqnarray}
\end{widetext}
 and
 \begin{widetext}
\begin{eqnarray}
\begin{split}
\mathcal{L}_{\,s}\rho =&
-\frac{1}{2}(\O{+}{}+\O{+}{12})
\big[(A_{ss}+A_{\e\e})\rho+\rho(A_{ss}+A_{\e\e})
-2(A_{s\e}+A_{\g s})\rho(A_{\e s}+A_{s\g})\big]
\\ &
-\frac{1}{2}(\O{-}{}+\O{-}{12})
\big[(A_{\e\e}-A_{ss})\rho+\rho(A_{\e\e}-A_{ss})
-2(A_{s\e}\rho A_{\e s}-A_{\g s}\rho A_{s\g})\big]
+i(\L{+}{}+\L{+}{12})(A_{s\e}\rho A_{s\g}-A_{\g s}\rho A_{\e s}),
\end{split}
\end{eqnarray}
\end{widetext}
respectively. It is clear that $\mathcal{L}_{\,s}\rho$ describes passage through the symmetric state $|s\rangle$, or the symmetric channel ($|\e\rangle-|s\rangle-|\g\rangle$), while $\mathcal{L}_{\,a}\rho$ describes the antisymmetric channel ($|\e\rangle-|a\rangle-|\g\rangle$). By writing (\ref{ME4}) in the molecular basis and obtaining (\ref{ME5}), we are then able to study these channels, which are represented in Figure \ref{Figura5}. 
\begin{figure}[ht]
\centering\includegraphics[width=0.45\columnwidth]{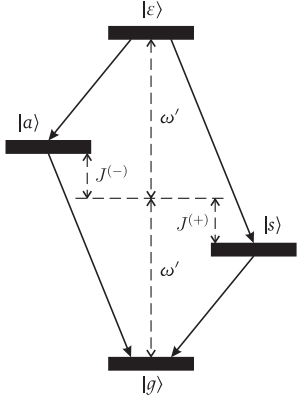}
\caption{ Collective states of the Hamiltonian (\protect{\ref{HS4t}}). The energies of the symmetric and antisymmetric states are shifted asymmetrically by the ``dressed'' coupling functions $J^{(\pm)}$ given by (\protect{\ref{cf}}) which are dependent on the XY coupling strength as well as on the relative atomic distance. The angular frequency $\omega'$ is given by (\protect{\ref{wf}}).
The sequence $|\e\rangle-|a\rangle-|\g\rangle$ represents the antisymmetric channel while the sequence $|\e\rangle-|s\rangle-|\g\rangle$ represents the symmetric channel.}
\label{Figura5}
\end{figure}

\begin{figure}[h]
\centering\includegraphics[width=0.8\columnwidth]{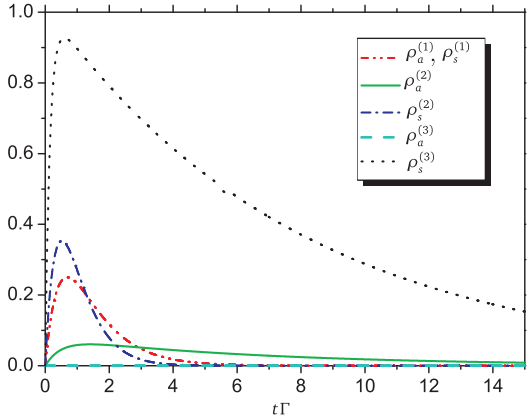}
\caption{Probability of the system initially prepared in $|\e\rangle$ to be found in the symmetric state $\rho^{(i)}_s$ or in the antisymmetric state $\rho^{(i)}_{a}$ for models (\protect{\ref{ME1}}), (\protect{\ref{ME2}}), and (\protect{\ref{ME4}}), denoted by $\rho^{(1)}$, $\rho^{(2)}$, and $\rho^{(3)}$, respectively. We are considering $r_{12}/\lambda=0.2$ and $J/\omega_0=0.6$.} 
\label{populacao1}
\end{figure}

From now on, calculations made with master equations (\ref{ME1}), (\ref{ME2}), and (\ref{ME4}) will be denoted $\rho^{(1)}$, $\rho^{(2)}$, and $\rho^{(3)}$, respectively. Atoms initially prepared in the excited state $|\e\rangle$ will decay to the ground state $|\g\rangle$ through both channels, but with different rates. The study of these decay channels will serve well to highlight the different behaviors predicted by (\ref{ME1}), (\ref{ME2}), and (\ref{ME4}). Figure \ref{populacao1} shows the populations $\rho_s$ of level $|s\rangle$ and $\rho_a$ of level $|a\rangle$, given the system was initially prepared in the excited state $|e\rangle$. In these plots, we used $r_{12}/\lambda=0.2$ and $J/\omega_0=0.6$. The probabilities are shown as functions of $\Gamma t$, where $\Gamma=|{\bf d}|^2\omega_0^3/(3\pi c^3\hbar\epsilon_0)$ is the atomic decay constant in the absence of another atom. 

In this illustrative case, i.e for $r_{12}/\lambda=0.2$ and $J/\omega_0=0.6$, which are very reasonable values, there is no difference between the populations of the symmetric and antisymmetric states according to (\ref{ME1}). For (\ref{ME2}), the symmetric channel is more accessed than the antisymmetric one. For the generalised master equation (\ref{ME4}), the result is quite remarkable. The system practically decays by using only the symmetric channel. The antisymmetric channel is very low populated for these parameters.

\section{Fluorescence Spectrum}\label{F}
In the previous section, we showed that the dynamics governed by the generalized master equation (\ref{ME4}) can be quite different then the one predicted by phenomenological models. Now, we also investigate spectrum of the system. Although mismatches between master equations (\ref{ME1}), (\ref{ME2}), and (\ref{ME4}) are not so drastic in this case, there are still differences which could be measured, especially by varying $J$ externally. To be more specific, we will study the fluorescence spectrum of this atomic system. This quantity measures the number of photons emitted by the atoms into the vacuum field modes as a function of the frequency of the modes \cite{ficek}. 

The fluorescence spectrum is evaluated as the real part of the Fourier transform of the two-time first-order correlation function $\langle E^{(-)}({\bf R},t)E^{(+)}({\bf R,t+\tau})\rangle$ for the positive $E^{(+)}$ and negative $E^{(-)}$ frequency components of the electric field operator at the position ${\bf R}$ of the detector \cite{ficek}. We can write the steady-state fluorescence spectrum as \cite{ficek}
\begin{eqnarray}
S(\omega)={\rm Re}\int_0^\infty{\rm d}\tau
\lim_{t\rightarrow\infty}\sum_{ij}\O{-}{ij\omega_0}
\langle\sigma_i^+(t)\sigma_j^-(t+\tau)\rangle e^{i\omega\tau}.
\end{eqnarray}
The fluorescence spectrum can be decomposed into coherent and incoherent components. The incoherent component of the spectrum $S_I(\omega)$ is obtained by subtracting the coherent component from the fluorescence spectrum as
\begin{eqnarray}\label{is}
S_I(\omega)&=&{\rm Re}\int_0^\infty{\rm d}\tau
\lim_{t\rightarrow\infty}\sum_{ij}\O{-}{ij\omega_0}[
\langle\sigma_i^+(t)\sigma_j^-(t+\tau)\rangle
\nonumber\\ && -\langle\sigma_i^+(t)\rangle\langle\sigma_j^-(t+\tau)\rangle] e^{i\omega\tau}.
\end{eqnarray}
Under the presence of the driving field, the system Hamiltonian becomes 
$H_{\mathcal F}=H_S^{(4)}+H_L,$
with
\begin{eqnarray}\label{aiai}
H_L&=&\frac{i\hbar}{2\sqrt{2}}\sum_{i=1}^{2}\big[
\Omega_L({\bf r}_i)\sigma_i^- e^{i\omega_L t}-{\rm H.c.}\big]\nonumber\\ &=&\frac{i\hbar}{4}\big[
(\Omega_L({\bf r}_1)+\Omega_L({\bf r}_2))(A_{s\e}+A_{\g s})e^{i\omega_L t}
\nonumber\\ &&+
(\Omega_L({\bf r}_2)-\Omega_L({\bf r}_1))(A_{a\e}-A_{\g a})
e^{i\omega_L t}
-{\rm H.c.}\big],
\end{eqnarray}
where $\Omega_L({\bf r}_i)$ is essentially the Rabi frequency for atom $i$. In the specific case of $\Omega_L({\bf r}_2)=\Omega_L({\bf r}_1)$, which will be considered from now on, it is clear from (\ref{aiai}) that the driving laser field can only interact with the symmetric channel. In this case, and by moving to an interaction picture defined by $\widetilde{H}_L=\exp(iH_S^{(4)}t/\hbar)H_L\exp(-iH_S^{(4)}t/\hbar)$, one obtains 
\begin{eqnarray} \label{HLRWA}
\widetilde{H}_L&=&\frac{i\hbar\Omega_L}{2}[
A_{se}e^{-i(\omega'+J^{(+)}-\omega_L)t}-A_{es}e^{+i(\omega'+J^{(+)}-\omega_L)t}
\nonumber\\ &&+A_{gs}e^{-i(\omega'-J^{(+)}-\omega_L)t}-A_{sg}e^{+i(\omega'-J^{(+)}-\omega_L)t}
],
\end{eqnarray}
where $\Omega_L\equiv\Omega_L({\bf r}_2)=\Omega_L({\bf r}_1)$. Essentially, (\ref{s1}) and (\ref{hs2}) transforms in the same way as (\ref{aiai}) but with different frequencies. 

\begin{figure}[h]
\centering\includegraphics[width=0.9\columnwidth]{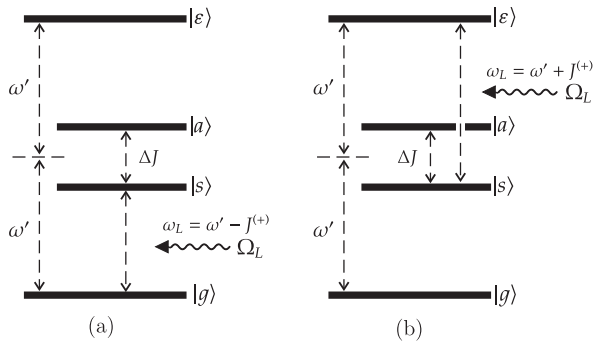}\vspace{-0.3cm}
\caption{Energy diagram of the system. I the part (a) the laser frequency is $\omega_L=\omega'-J^{(+)}$. In part (b) the frequency of the laser is $\omega_L=\omega'+J^{(+)}$. In this picture $\Delta J\equiv J^{(-)}-J^{(+)}$.}  
\label{2cases}
\end{figure}

\begin{figure}[ht]
\centering\includegraphics[width=0.8\columnwidth]{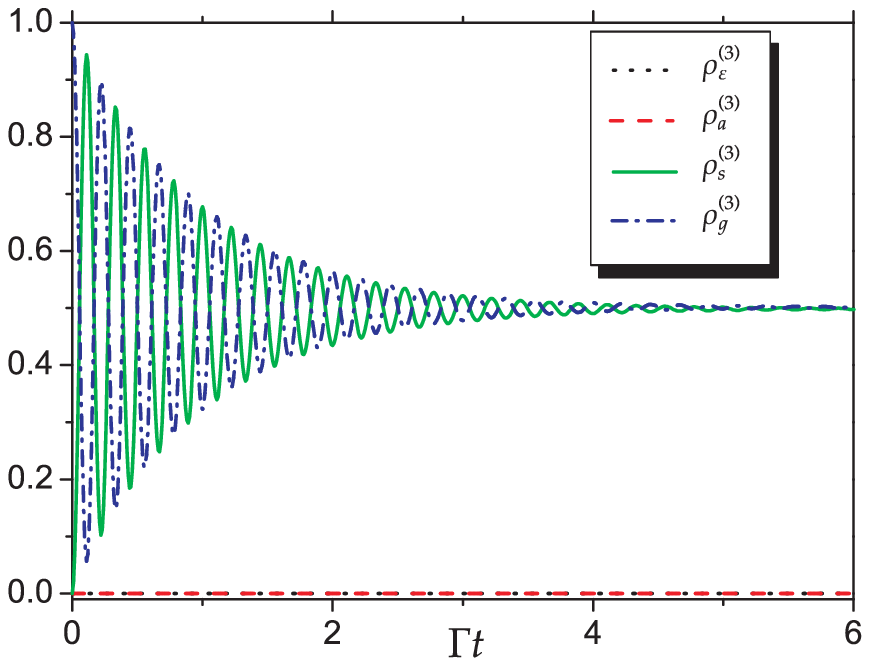}
\caption{Probability of the system initially prepared in $|\g\rangle$ to be found in the ground $\rho^{(3)}_\g$,  excited $\rho^{(3)}_\e$, symmetrical $\rho^{(3)}_s$ or antisymmetric $\rho^{(3)}_s$ states. The driving field frequency is $\omega_L=\omega'-J^{(+)}$. In this plot, we used $J/\omega_0=0.1$, $\Omega_L/\omega_0=10$ and $r_{12}/\lambda=0.2$.}
\label{populcao3}
\end{figure}

\begin{figure}[ht]
\centering\includegraphics[width=0.8\columnwidth]{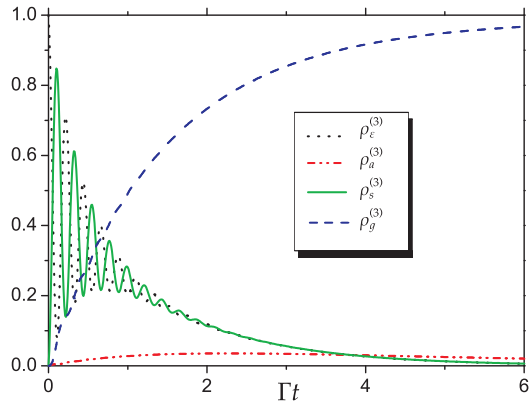}
\caption{ Probability of the system initially prepared in $|\e\rangle$ to be found in the ground $\rho^{(3)}_\g$,  excited $\rho^{(3)}_\e$, symmetrical $\rho^{(3)}_s$ or antisymmetric $\rho^{(3)}_s$ states. The driving field frequency is $\omega_L=\omega'+J^{(+)}$. In this plot, we used $J/\omega_0=0.1$, $\Omega_L/\omega_0=10$ and $r_{12}/\lambda=0.2$.}
\label{populcao2}
\end{figure}

Let us start by considering $\omega_L=\omega'-J^{(+)}$, which is the case depicted in the Figure (\ref{2cases}) panel (a). From (\ref{HLRWA}), we can see that terms like $A_{se}$ and $A_{es}$ can, in principle, be ignored by means of the rotating wave approximation (RWA). If this is really the case, for the system initially prepared in $|\g\rangle$, a laser could only (or with great probability) induce transitions between the symmetrical and ground states. From Figure \ref{populcao3}, we can see that this is indeed the case. As a direct consequence of this coupling between $|g\rangle$ and $|s\rangle$, the system is driven to a steady state which is different from $|g\rangle$ allowing us to obtain a spectrum. For $\omega_L=\omega'+J^{(+)}$, the situation is depicted in Figure \ref{2cases} panel (b). Now, similar arguments used before show that $A_{gs}$ and $A_{sg}$ will be not important in (\ref{HLRWA}). So, regardless the initial state preparation, if the system reaches the ground state, the laser will not be able to move it from there. Consequently, there is no steady state other than $|g\rangle$. This situation is presented in the Figure (\ref{populcao2}), where the system is considered to be initially prepared in the excited state $|\e\rangle$.

So, for the case of interest, where $\omega_L\approx \omega'-J^{(+)}$, we can write the effective Hamiltonian 
\begin{eqnarray}\label{HS3}
H_{\mathcal F}^{{\rm{eff}}}&=&\frac{\delta' \hbar}{2}\sigma^Z_{sg}+i\hbar\frac{\Omega_L}{2}( A_{\g s}e^{i\omega_L t}- A_{s\g}e^{-i\omega_L t}),
\end{eqnarray}
where we defined the detuning $\delta'=\omega'-J^{(+)}$ and $\sigma_{sg}^Z=A_{ss}-A_{\g\g}$. We can see that the system behaves as a two-level system being driven by a laser field with Rabi frequency $\Omega_L$. As mentioned before, this will also be the case with (\ref{s1}) and (\ref{hs2}) but with different detunings involved.

\begin{figure}[ht]
\centering\includegraphics[width=\columnwidth]{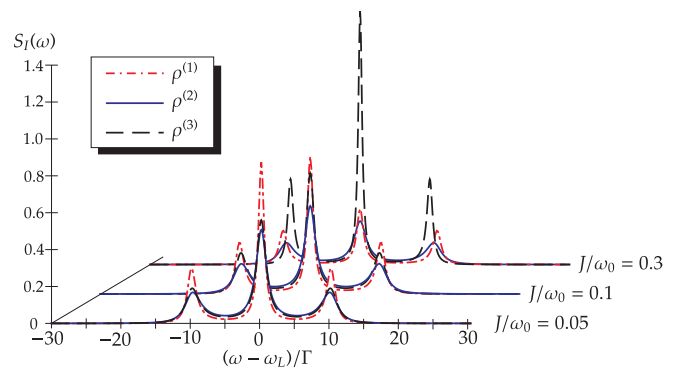}
\caption{Incoherent fluorescent spectrum for different values of the coupling $J$. In this figure, we consider $r_{12}/\lambda=0.2$ and $\Omega_L/\omega_0=10$ and $\omega_L=\delta'-J^{(+)}$.}
\label{espectro}
\end{figure}

\begin{figure}[ht]
\centering\includegraphics[width=0.8\columnwidth]{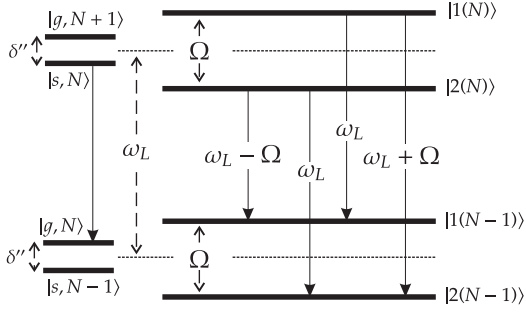} 
\caption{Splitting of the states by the Stark effect. Allowed spontaneous transitions between uncoupled states (on the left)  and dressed states (on the right). In this picture, $\delta''=\delta'-\omega_L$ and $\Omega=\sqrt{(\delta'')^2+\Omega_L^2}$ and $|N\rangle$ is the Fock state for the quantized laser field \protect{\cite{cohen}}.}
\label{mollowtriplet}
\end{figure}

In Figure (\ref{espectro}), it is shown the incoherent fluorescent spectrum (\ref{is}) calculated with master equations (\ref{ME1}),  (\ref{ME2}), and (\ref{ME4}), again identified by $\rho^{(1)}$, $\rho^{(2)}$, and $\rho^{(3)}$, respectively.  In all cases, we obtain the expected Mollow or Stark triplet \cite{PhysRev.188.1969}. This triplet is depicted in Figure (\ref{mollowtriplet}). The physical explanation for the formation of this triplet is well known \cite{cohen}, and it is given by using the dressed atom picture and treating the laser quantum mechanically. From Figure (\ref{espectro}), we can see that (\ref{ME1}), (\ref{ME2}), and (\ref{ME4}) lead to different spectra and that these differences can still be detected by varying $J$. An experiment can, in principle, be conceived to measure the intensity of the central peak as a function of $J$, in some arbitrary units, and then check which model fits better. Given that our generalized master equation (\ref{ME4}) is based on a microscopic model which takes the $J$ coupling from the start, we believe it to give more accurate answers than (\ref{ME1}) or (\ref{ME2}).

\section{Conclusions}\label{S}
We have derived a master equation describing two dissipative XY-coupled qubits taking into account the qubit-qubit interaction in the microscopic model. We then applied this development to atomic physics where we discussed the dynamics of populations and the fluorescence spectrum. We highlighted instances where the phenomenological models, which consider the qubit-qubit coupling only \textit{a posteriori}, lead to different predictions when compared to the microscopic model developed here. In particular, we studied the populations of the eigenstates of the XY Hamiltonian and the fluorescence spectrum. 

It would be interesting to see similar investigations in other physical setups which are traditionally described by using phenomenological master equations. One could, for instance, study the microscopic derivation of master equations for the many generalizations of the Jaynes-Cummings model, including multilevel atoms \cite{3}, 
external fields \cite{4}, multi-atom configurations \cite{5}, and
multi-photon transitions \cite{6}, just to mention a few examples. In all these cases, one could perform similar studies as the one developed here and in \cite{sabrina}, trying to obtain master equations obtained with microscopic models which already include the generalized Jaynes-Cummings Hamiltonians from the beginning. 

\begin{acknowledgments}
J.P.S. acknowledges Fundação de Amparo a Pesquisa do
Estado de São Paulo (FAPESP) Grant No. 2011/09258-5,
Brazil. F.L.S. acknowledges partial support from CNPq (Grant
No. 308948/2011-4) and the Brazilian National Institute of
Science and Technology of Quantum Information (INCT-IQ). We also thank Mauro Paternostro for insightful discussions.
\end{acknowledgments}

\appendix

\section{} \label{app1}
In this Appendix, we present the Hamiltonians, Liouvillians and coefficients necessary to work with the phenomenological master equations (\ref{ME1}), (\ref{ME2}) and with the generalized master equation (\ref{fi}). For (\ref{ME1}), one needs
\begin{eqnarray}
H_S^{(1)}&=&\frac{1}{2}\hbar(\omega_1+
\L{-}{\omega_1})\sigma_1^{z}+\frac{1}{2}\hbar(\omega_2+
\L{-}{\omega_2})\sigma_2^{z}\nonumber\\&&-\hbar J(\sigma_1^+\sigma_{2}^-+{\rm H.c.}),\label{s1}
\end{eqnarray}
and
\begin{eqnarray}
\mathcal{L}_i(\rho)=\frac{
\O{-}{\omega_i}
}{2}\big(2
\sigma_i^{-}\rho\sigma_i^{+}-\{\sigma_i^{+}\sigma_i^{-},\rho\}\big)\label{l1},
\end{eqnarray}
with spontaneous decay rate $\O{-}{\omega_i}$ and Lamb shift $\L{-}{\omega_i}$ given by
\begin{eqnarray}
\label{Oi}
\O{-}{\omega_i}=2\pi\sum_\lambda\int{\rm d}{\bf k}g({\omega_k})|\kappa_{i;{\bf k},\lambda}^{}|^2\delta(\omega_k-\omega_i),
\end{eqnarray}
and
\begin{eqnarray}
\label{Li}
\L{-}{\omega_i}=\sum_\lambda\int{\rm d}{\bf k}\frac{g(\omega_k)|\kappa_{i;{\bf k},\lambda}^{}|^2}{\omega_i-kc},&
\end{eqnarray}
respectively.
In these expressions, $g({\omega_k})$ is a density of states,  $\kappa^{}_{i;{\bf k},\lambda}$ is a coupling constant (for the moment unspecified), and  $i=\{1,2\}$. For the phenomenological master equation (\ref{ME2}), one also needs
\begin{eqnarray}
H_S^{(2)}&=&
\frac{1}{2}\hbar(\omega_1+
\L{-}{\omega_1})\sigma_1^{z}+\frac{1}{2}\hbar(\omega_2+
\L{-}{\omega_2})\sigma_2^{z}\nonumber\\&&-\hbar\big[(J-\L{-}{12\omega_0})\sigma_1^+\sigma_2^-+{\rm H.c.}\big]
,\label{hs2}
\end{eqnarray}
and
\begin{eqnarray}
\mathcal{L}_{ij}(\rho)=\frac{
\O{-}{ij\omega_0}}{2}\big(2
\sigma_i^{-}\rho\sigma_j^{+}-\{\sigma_i^{+}\sigma_j^{-},\rho\}\big),\label{l12}
\end{eqnarray}
where the collective parameters are given by
\begin{eqnarray}
\label{Oij}
\O{$\pm$}{ij\mu}=2\pi\sum_\lambda\int{\rm d}{\bf k}g(\omega_k)\kappa^{}_{i;{\bf k},\lambda}\kappa^*_{j;{\bf 
k},\lambda}\delta(kc\pm\mu),\end{eqnarray}
and
\begin{eqnarray}
\label{Lij}
\L{$\pm$}{ij\mu}&=&\sum_\lambda\int{\rm d}{\bf k}\frac{g(\omega_k)\kappa_{i;{\bf k},\lambda}^{}\kappa^*_{j;{\bf 
k},\lambda}}{\mu\pm kc},
\end{eqnarray}
respectively.
In (\ref{hs2}) and (\ref{l12}), we use (\ref{Oij}) and (\ref{Lij}) with $\mu=\omega_0$ where $\omega_0\equiv(\omega_1+\omega_2)/2$. In the following, we keep using (\ref{Oij}) and (\ref{Lij}), but with $\mu$ assuming other forms depending on the model.

In order to use (\ref{fi}), one needs the coefficients $\mathcal{A}_{ij}$, $\mathcal{B}_{ij}$, $\mathcal{C}_{ij}$, $\mathcal{D}_{ij}$, $\mathcal{E}_{ij}$ and $\mathcal{F}_{ij}$. For this, it is convenient to first define 
\begin{eqnarray}
\O{$\pm$}{i\mu}&\equiv&\O{$\pm$}{ii\mu},\label{def1}\\
 \L{$\pm$}{i\mu}&\equiv&\L{$\pm$}{ii\mu}, \label{def2}
\end{eqnarray}
and also
\begin{eqnarray}
\L{$\pm$}{i}&=&(\L{-}{i\alpha}\pm\L{+}{i\beta}),\label{def3}
\\
\O{$\pm$}{i}&=&\frac{1}{2}(\O{-}{i\alpha}\pm\O{+}{i\beta}),\label{def4}
\\
\L{$\pm$}{12}&=&(\L{-}{12\alpha}\pm\L{+}{12\beta}),\label{def5}
\\
\O{$\pm$}{12}&=&\frac{1}{2}(\O{-}{12\alpha}\pm\O{+}{12\beta}),\label{def6}
\end{eqnarray}
with $\mu=\{\alpha,\beta\}$, where $\alpha$ and $\beta$ are given by (\ref{alpha}) and (\ref{beta}), respectively. Now, we can  present the definitions of $\mathcal{A}_{ij}$, $\mathcal{B}_{ij}$, $\mathcal{C}_{ij}$, $\mathcal{D}_{ij}$, $\mathcal{E}_{ij}$ and $\mathcal{F}_{ij}$ 

\begin{eqnarray}\label{coef}
\mathcal{A}_{11}&=&
\gamma\O{-}{1\alpha}+\delta\O{+}{1\beta},\nonumber \\
\mathcal{A}_{22}&=&
\delta\O{-}{2\alpha}+\gamma\O{+}{2\beta},\nonumber \\
\mathcal{A}_{12}&=&
\cO{+}{12}
+i\frac{(\omega_1-\omega_2)}{\Delta}
\cL{+}{12}, \nonumber \\
\mathcal{A}_{21}&=&\mathcal{A}_{12}^*,\nonumber \\
\mathcal{B}_{11}&=&\mathcal{G}_5 |\theta_1(t)|^2+\mathcal{G}_6 |\theta_2(t)|^2\nonumber \\
\mathcal{B}_{22}&=&\mathcal{G}_5 |\theta_2(t)|^2+\mathcal{G}_6 |\phi_1(t)|^2\nonumber \\
\mathcal{B}_{21}&=&
i(\delta\cL{+}{12\beta}-\gamma\cL{-}{12\alpha})
-\frac{1}{2}(\delta\cO{+}{12\beta}
+\gamma\cO{-}{12\alpha})
+\frac{J}{\Delta}\Big[
\O{-}{2}+i\L{+}{2}\Big],
\nonumber \\
\mathcal{B}_{12}&=& 
i(\gamma\L{+}{12\beta}-\delta\L{-}{12\alpha})-
\frac{1}{2}(\delta\O{-}{12\alpha}+\gamma\O{+}{12\beta})+\frac{J}{\Delta}\Big[
\O{-}{1}+i\L{+}{1}\Big],\nonumber \\
\mathcal{C}_{12}&=&\frac{J}{\Delta}
\Big[\O{-}{1}+i\L{+}{1}\Big],\nonumber\\
\mathcal{C}_{21}&=&\frac{J}{\Delta}
\Big[\O{-}{2}+i\L{+}{2}\Big],\nonumber \\
\mathcal{D}_{12}&=&\frac{J}{\Delta}
\Big[\cO{-}{12}+i\cL{+}{12}\Big], \nonumber\\
\mathcal{D}_{21}&=&\frac{J}{\Delta}
\Big[\O{-}{12}+i\L{+}{12}\Big],\nonumber \\
\mathcal{E}_{12}&=&
\mathcal{G}_1\theta_1^*(t)\theta_2(t)+
\mathcal{G}_2|\theta_1(t)|^2+
\mathcal{G}_3|\theta_2(t)|^2,
 \nonumber\\
\mathcal{E}_{21}&=&
\mathcal{G}_1\phi_1(t)\theta_2^*(t)+
\mathcal{G}_2|\theta_2(t)|^2+
\mathcal{G}_3|\phi_1(t)|^2,\nonumber \\
\mathcal{F}_{12}&=&
\mathcal{G}_4\theta_1(t)\theta_2^*(t),
\nonumber\\
\mathcal{F}_{21}&=&
\mathcal{G}_4\phi_1^*(t)\theta_2(t),
\end{eqnarray}
where the auxiliary functions $\mathcal{G}_i$ are giving by
\begin{eqnarray}
\mathcal{G}_1&=&
\theta_2(t)\Big(\gamma e^{it\beta}\Ph{-}{2\beta}+\delta e^{-it\alpha}\Ph{+}{2\alpha}\Big)
-\theta_1(t)\Big(\gamma e^{it\beta}\Ph{-}{12\beta}+\delta e^{-it\alpha}\Ph{+}{12\alpha}\Big)
\nonumber\\ &&+\frac{J}{\Delta}\Big[
\theta_1(t)(e^{-it\alpha}\Ph{+}{1\alpha}-e^{it\beta}\Ph{-}{1\beta})-\theta_2(t)(e^{-it\alpha}\cPh{-}{12\alpha}-e^{it\beta}\cPh{+}{12\beta})
\Big]\nonumber\\
\mathcal{G}_2&=&
-\theta_1(t)\frac{J}{\Delta}\Big(
e^{-it\alpha}\Ph{+}{12\alpha}
-e^{it\beta}\Ph{-}{12\beta}\Big)-\theta_2(t)\Big(\delta e^{i\beta t}\cPh{+}{12\beta}
+\gamma e^{-i\alpha t}\cPh{-}{12\alpha}\Big)\nonumber\\
\mathcal{G}_3&=&
-\phi_1(t)\frac{J}{\Delta}\Big(e^{-it\alpha}\cPh{-}{12\alpha}-e^{it\beta}\cPh{+}{12\beta}\Big)
-\theta_2(t)\Big(\gamma e^{i\beta t}\Ph{-}{12\beta}+\delta e^{-i\alpha t}\Ph{+}{12\alpha}\Big)\nonumber\\
\mathcal{G}_4&=&
\theta_2(t)\Big(\delta e^{it\beta}\Ph{-}{1\beta}+\gamma e^{-it\alpha}\Ph{+}{1\alpha}\Big)
-\phi_1(t)\Big(\delta e^{it\beta}\cPh{+}{12\beta}+\gamma e^{-it\alpha}\cPh{-}{12\alpha}\Big)
\nonumber\\
&&+\frac{J}{\Delta}\Big[
\phi_1(t)(e^{-it\alpha}\Ph{+}{2\alpha}-e^{it\beta}\Ph{-}{2\beta})
-\theta_2(t)(e^{-it\alpha}\Ph{+}{12\alpha}-e^{it\beta}\Ph{-}{12\beta})
\Big]\nonumber\\
\mathcal{G}_5&=&
-\theta_1(t)\Big(\delta e^{i\beta t}\Ph{-}{1\beta}+\gamma e^{-it\alpha}\Ph{+}{1\alpha}\Big)
-\theta_2(t)\frac{J}{\Delta}\Big(e^{-i\alpha t}\Ph{+}{2\alpha}- e^{it\beta}\Ph{-}{2\beta}\Big) 
\nonumber\\
\mathcal{G}_6&=&
-\phi_1(t)\Big(\gamma e^{i\beta t}\Ph{-}{2\beta}+\delta e^{-it\alpha}\Ph{+}{2\alpha}\Big)
-\theta_2(t)\frac{J}{\Delta}\Big(
e^{-i\alpha t}\Ph{+}{1\alpha}-e^{it\beta}\Ph{-}{1\beta}\Big)\nonumber\\ 
\end{eqnarray}
with
\begin{eqnarray}
\Ph{$\pm$}{i\alpha}
&=&
\frac{1}{2}\O{-}{i\alpha}
\pm i\L{-}{i\alpha}, 
\nonumber\\
\Ph{$\pm$}{i\beta}
&=&
\frac{1}{2}\O{+}{i\beta}
\pm i\L{+}{i\beta}, 
\nonumber\\
\Ph{$\pm$}{12\alpha}
&=&
\frac{1}{2}\O{-}{12\alpha}
\pm i\L{-}{12\alpha}
\nonumber\\
\Ph{$\pm$}{12\beta}
&=&
\frac{1}{2}\O{+}{12\beta}
\pm i\L{+}{12\beta}
\end{eqnarray} 
\end{document}